\documentclass{jpsj-suppl}
\usepackage{txfonts} 
\usepackage{graphicx,color}

\title{Reflective terahertz time-domain spectroscopy measurement on the stripe-ordered superconductor La$_{1.84-y}$Nd$_y$Sr$_{0.16}$CuO$_4$}

\author{Tyler \textsc{Miyake}$^{1}$, Shinya Miyazaki$^1$, Yusuke Sakai$^1$, Kiyohisa Tanaka$^1$, Shigeki Miyasaka$^1$, Setsuko Tajima$^1$, and Masayoshi Tonouchi$^2 $ }

\inst{$^{1}$Graduate School of Science, Osaka University, 1-1 Machikaneyama-cho, Toyonaka, Osaka, Japan, 560-0043 \\
$^{2}$Institute of Laser Engineering, Osaka University, 2-6, Yamadaoka, Suita, Osaka, Japan, 565-0871}

\email{miyake@tsurugi.phys.sci.osaka-u.ac.jp}

\recdate{June 23, 2011}

\abst{
We measured reflectivity on static stripe-ordered La$_{1.84-y}$Nd$_y$Sr$_{0.16}$CuO$_4$ (LNSCO) $y$\,=\,0.1, 0.2, 0.3, and 0.4 by means of reflective terahertz time-domain spectroscopy (THz-TDS) with electric field polarized along $c$-axis, in which one can obtain lower frequency information than the conventional Fourier transform type spectrometer. 
Recently, two-dimensional superconducting (2DSC) property \cite{tajima,schafgans,li}, intra-layer perfect conductivity not accompanied with inter-layer superconducting behavior, was reported on LNSCO ($x$\,=0.15\,,\,$y$\,$\geq$\,0.2).
We observed the existence of Josephson plasma edge, which suggests that 2DSC is not realized in these compounds.}

\kword{high-$T\mathrm{_c}$, stripe order, reflective terahertz time-domain spectroscopy (THz-TDS)}

\begin{document}
\maketitle

\section{Introduction}
The "stripe order" is a periodic modulation of one-dimensional charge and spin densities seen in some hole-doped transition metal oxides.
It was observed by means of neutron scattering, X-ray diffraction and so on \cite{tranquada, hucker}. 
In hole-doped La-214 cuprates, such as La$_{2-x}$Ba$_{x}$CuO$_4$ (LBCO) and La$_{1.84-y}$Nd$_y$Sr$_{0.16}$CuO$_4$ (LNSCO), represent this charge and magnetic order coexisting with superconductivity.

The static stripe order in LNSCO is triggered by the structural phase transition from low temperature orthorhombic (LTO) to the low temperature tetragonal (LTT) phase, whose transition temperature $T\mathrm{_{LT}}$ is higher than the superconducting transition temperature ($T\mathrm{_c}$) \cite{buchner, noda, singer, crawford}.
The structural transition temperature increases and the stability of the static stripe order is enhanced with increasing Nd concentration $y$, without changing hole concentration \cite{singer, crawford}. 
As the stripe order is stabilized, the superconducting order is disturbed and $T\mathrm{_c}$ becomes lower. 

Terahertz time-domain spectroscopy (THz-TDS) is a spectroscopic technique to explore low frequency excitations in solids, which has become available recently. 
Compared to conventional Fourier transform infrared spectroscopy (FT-IR), THz-TDS has several advantages. First, THz-TDS can detect extremely low frequency signal down to 0.1 THz (3.3\,cm$^{-1}$), which is not available in FT-IR. 
Secondly, since both amplitude and phase spectra can be obtained by computing the Fourier transforms of the measured time-domain wave forms of the THz radiation, one can derive complex optical functions without extrapolating the reflectivity spectra and resorting to the Kramers-Kronig analysis. 
Reflective-type THz-TDS system applied in this study is more useful than transmissive-type one especially for bulk superconductors, because the reflectivity reaches nearly 100\% below $T\mathrm{_c}$ and there is almost no signal in transmissive-type one. 

Recently, destruction of the inter-layer Josephson coupling with intra-layer superconductivity is reported on the static stripe ordered high-$T\mathrm{_c}$ compounds including LNSCO ($x$\,=\,0.15, $y$\,$\geq$\,0.2) \cite{tajima, li, schafgans}. 
This superconducting phase coherence only in the CuO$_2$ plane is regarded as "two-dimensional superconductivity (2DSC)" or "decoupled state".
However, because of the limited $\omega$-range in FT-IR spectroscopy, it is not clear whether a Josephson plasma completely disappears or not.

In this study, we re-examined reflectivity of the stripe ordered LNSCO using our reflective THz-TDS system. 
The aim of this study is to clarify whether the Josephson coupling survives in low frequency as the stripe order is stabilized and whether 2DSC states is realized or not.

\section{Experimental}
A schematic picture and detail of our reflective THz-TDS system is shown in ref. \cite{matsuoka}. 
For the emission and the detection of the pulsed THz radiation, we use a Ti:\,sapphire pulsed laser (wavelength 800\,nm, width 100\,fs, repetition rate 80\,MHz) and photoconductive (PC) dipole and bow-tie antennas, made by evaporating gold in a dipole- or bow-tie-like pattern on the low-temperature grown GaAs substrate. 
The polarized reflectivity measurements can be carried out using two wire-grid polarizers. 
The samples were set in a temperature-controllable cryostat (4.8\,K\,-\,300\,K) with a gold mirror used as a reference. 

LNSCO single crystals ($x$\,=\,0.16, $y$\,=\,0.1, 0.2, 0.3, and 0.4) were grown by a traveling solvent floating-zone (TSFZ) method. 
As-grown crystals were cut parallel to the $ac$-plane and polished to obtain the surfaces for the spectroscopy measurements. 
After polish, samples were annealed on the gold plate in O$_2$  atmosphere at 800$^\circ$C for 1 week and quenched to the room temperature to make the hole concentrations uniform.

\begin{figure}[tbh]
\begin{center}
\includegraphics[width=7cm]{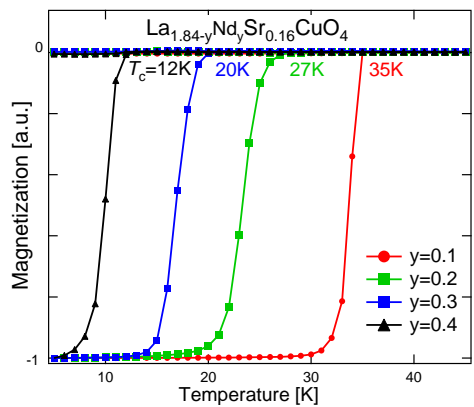}
\caption{Normalized magnetization curves of La$_{1.84-y}$Nd$_y$Sr$_{0.16}$CuO$_4$ with $y$\,=\,0.1\/(a), 0.2\,(b), 0.3\,(c), 0.4\,(d) after annealing. 
The measurements were operated under a weak magnetic field $H$\,=\,10\,Oe.} 
\label{f1}
\end{center}
\end{figure}

Superconducting transition temperatures of the samples were determined as $T\mathrm{_c}$\,=\,35\,K for $y$\,=\,0.1, 27\,K for $y$\,=\,0.2, 20\,K for $y$\,=\,0.3, and 12\,K for $y$\,=\,0.4, from the onset in the magnetization measurements (Fig. ~\ref{f1}). 

\section{Results and discussion}
Temperature dependence of the time-domain wave forms are shown in Fig. 2. 
Fig. 3 shows the reflectivity spectra obtained by taking Fourier transform of the measured time-domain wave forms. 
The fact that the edge-like structure observed in superconducting states for $y$\,=\,0.1, 0.2, and 0.3 disappears above $T\mathrm{_c}$ indicates we successfully observed the Josephson plasma edge. 
(For $y$\,=\,0.4, the tail of the Josephson plasma edge seems to be observed). 
The $y$-dependence of reflectivity shows that the Josephson plasma edge shifts to the lower frequency as $y$ increases. 
This behavior is associated with the decrease in the superfluid density $\rho \mathrm{_s}$ along $c$-axis ($\omega \mathrm{_p} \propto \sqrt{\rho \mathrm{_s}}$).

It is clear that the Josephson plasma edge can be seen even in the heavily Nd-doped sample $y$\,=\,0.3 (and $y$\,=\,0.4) in the LTT-phase, where the stripe order is static.
These results suggest that the inter-layer Josephson coupling can coexist with static stripe order at least in optimally doped compounds and previous reports could not observe  the Josephson plasma edge because of the detection limitation of the low frequency signal \cite{tajima, schafgans}.

We also found anomaly in the reflectivity spectra. 
In $y$\,=\,0.2 and 0.3, reflectivity does not go to 100\% below the Josephson plasma frequency and shows absorption structure in the low frequency below 0.5\,THz. 
Since all the measurements were performed under $T\mathrm{_{LT}}$ ($T\mathrm{_{LT}}=45\mathrm{K}$,\,62K,\,75K, for $y=0.2$,\,0.3,\,and 0.4, respectively), it may be related to the static stripe order. 
It should be also noted that this structure can be observed also in the normal state and may come from the noise during the Fourier transfer procedure of the time-domain wave form, however, such structure was never observed in La$_{2-x}$Sr$_{x}$CuO$_4$ (LSCO) $x$\,=\,0.1 \cite{matsuoka}, whose Josephson plasma frequency is similar to LNSCO $y$\,=\,0.3. 
To clarify the origin of this structure, it is necessary to perform the measurement at higher temperature across LTO-LTT structural phase transition temperature \cite{buchner, noda, singer, crawford}, and proper theory is also needed.

\begin{figure}[tbh]
\begin{center}
\includegraphics[width=14cm]{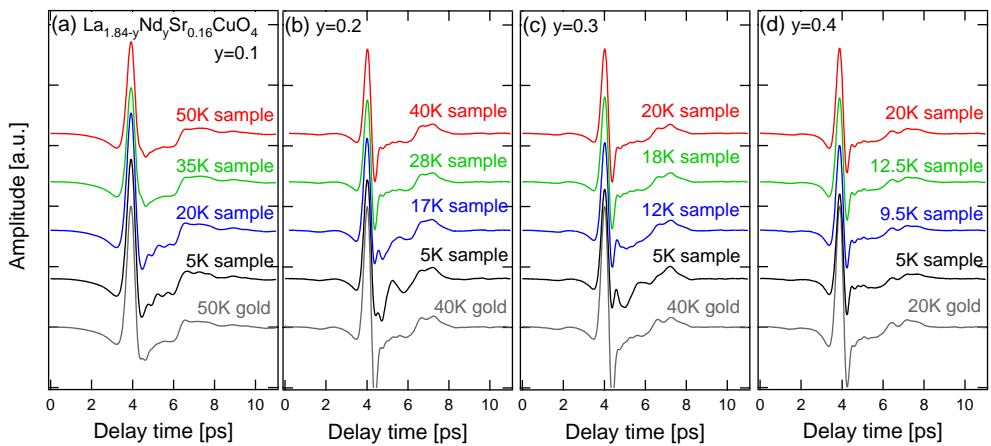}
\caption{Time-domain wave forms of the reflected THz pulsed wave of La$_{1.84-y}$Nd$_y$Sr$_{0.16}$CuO$_4$ with $y$\,=\,0.1\/(a), 0.2\,(b), 0.3\,(c), 0.4\,(d) and gold mirror as the reference.} 
\label{f2}
\end{center}
\end{figure}

\begin{figure}[tbh]
\begin{center}
\includegraphics[width=14cm]{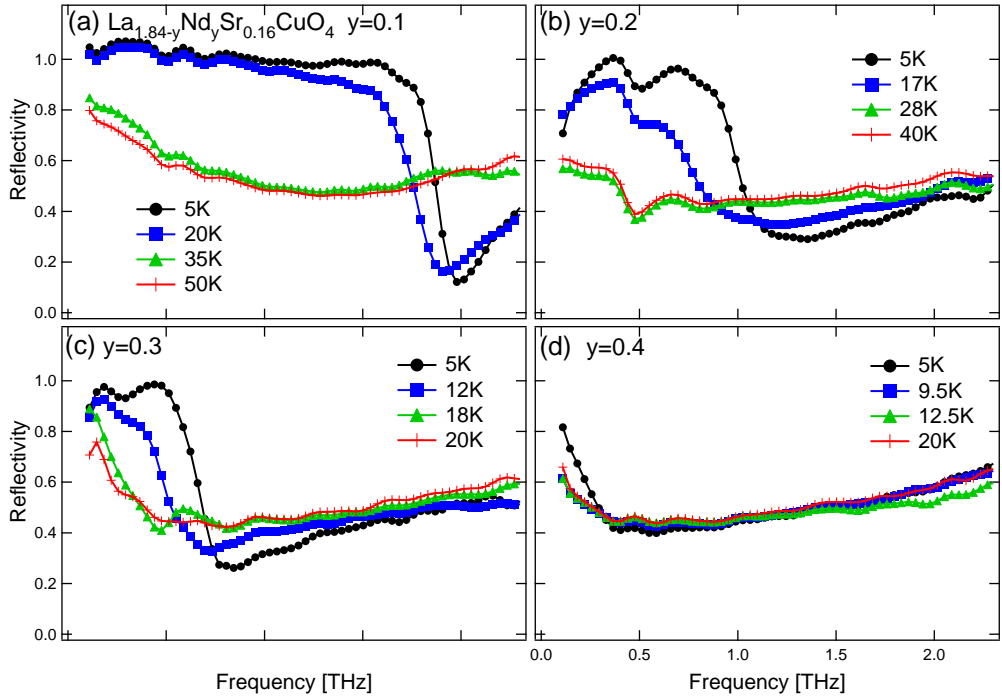}
\caption{Reflectivity spectra of La$_{1.84-y}$Nd$_y$Sr$_{0.16}$CuO$_4$ with $y$\,=\,0.1\/(a), 0.2\,(b), 0.3\,(c), 0.4\,(d).}
\label{f3}
\end{center}
\end{figure}

\section{Conclusion}
In this study, we performed the THz-TDS measurements on the static stripe-ordered LNSCO. 
We successfully observed the Josephson plasma edge in the low frequency region, which can not be measured in a conventional Fourier transform type spectrometer. 
This result suggests the coexisting of the interlayer phase coherence and the static stripe order, indicating 2DSC is not realized in LNSCO $x$\,=\,0.16, $y$\,$\geq$0.2.
We also observe the anomalous absorption structure in reflectivity spectra which may be related to the stripe order. 
To clarify the origin of this structure, further studies including measurements on other stripe compounds are needed.

\section*{Acknowledge}
We would like to thank T. Taniguchi, K. Makisaka, and T. Kondoh (Osaka Univ.) for their support in operating crystal growth by TSFZ method, and N. Matsuura (Institute for Material Research, Tohoku Univ.) for his suggestion about the crystal growth techniques.


\begin{thebibliography}{9}
\bibitem{tajima} S. Tajima, T. Noda, H. Eisaki, and S. Uchida: Phys. Rev. Lett. {\bf 86} (2001) 500. 
\bibitem{schafgans} A. A. Schafgans, C. C. Homes, G. D. Gu, Seiki Komiya, Yoichi Ando, and D. N. Basov: Phys. Rev. B {\bf 82} (2010) 100505(R).
\bibitem{li} Q. Li, M. H\"{u}cker, G.D. Gu, A. M. Tsvelik, and J. M. Tranquada: Phys. Rev. Lett. {\bf 99} (2007) 067001.
\bibitem{tranquada} J. M. Tranquada, B. J. Sternlieb, J. D. Axe, Y. Nakamura, and S. Uchida: Nature, {\bf 375} (1995) 561.
\bibitem{hucker} M. H\"{u}cker,M. v. Zimmermann, G. D. Gu, Z. J. Xu, J. S. Wen, Guangyong Xu, H. J. Kang, A. Zheludev, and J. M. Tranquada: Phys. Rev. B {\bf 83} (2011) 104506.
\bibitem{buchner} B.B\"{u}chner, M. Breuer, A. Freimuth, and A. P. Kampf: Phys. Rev. Lett. {\bf 73} (1994) 1841.
\bibitem{noda} T. Noda, H. Eisaki, S. Uchida: Science, {\bf 286} (1999) 265.
\bibitem{crawford} M. K. Crawford, R. L. Harlow, E. M. McCarron, W. E. Farneth, J. D. Axe, H. Chou, and Q. Huang: Phys. Rev. Lett. {\bf 44} (1991) 7749.
\bibitem{singer} P. M. Singer, A. W. Hunt, A. F. Cederstr\"{o}m, and T. Imai: Phys. Rev. Lett. {\bf 60} (1999) 15345.
\bibitem{matsuoka} T. Matsuoka, T. Fujimoto, K. Tanaka, S. Miyasaka, S. Tajima, K. Fujii, M. Suzuki, and M. Tonouchi: Physica C {\bf 469} (2009) 982.

\end{thebibliography}
\end{document}